\documentclass[a4paper,twocolumn,superscriptaddress,11pt,accepted=2017-12-05]{quantumarticle}
\pdfoutput=1
\usepackage{amsmath}
\usepackage{hyperref}
\usepackage[numbers,sort&compress]{natbib}

\usepackage{epsfig}
\usepackage[dvipsnames]{xcolor}
\usepackage{color}
\usepackage{graphicx}
\usepackage{bm}
\providecommand{\openone}{\leavevmode\hbox{\small1\kern-3.8pt\normalsize1}}

\begin{document}
\title{Cavity assisted measurements of heat and work in optical lattices}

\author{Louis Villa}
\affiliation{Univ Lyon, Ens de Lyon, Univ Claude Bernard,
F-69342 Lyon, France}
\affiliation{Centre for Theoretical Atomic, Molecular and Optical Physics, School of Mathematics and Physics, Queen's University, Belfast BT7 1NN, United Kingdom}
\author{Gabriele De Chiara}
\email{g.dechiara@qub.ac.uk}
\affiliation{Centre for Theoretical Atomic, Molecular and Optical Physics, School of Mathematics and Physics, Queen's University, Belfast BT7 1NN, United Kingdom}

\date{\today }

\maketitle

\begin{abstract}
We propose a method to experimentally measure the internal energy of a system of ultracold atoms trapped in optical lattices by coupling them to the  fields of two optical cavities. We show that the tunnelling and self-interaction terms of the one-dimensional Bose-Hubbard Hamiltonian can be mapped to the field and photon number of each cavity, respectively. We compare the energy estimated using this method with numerical results obtained using the density matrix renormalisation group algorithm. Our method can be employed for the assessment of power and efficiency of thermal machines whose working substance is a strongly correlated many-body system.
\end{abstract}



\section{Introduction}
The precise measurement of energy of a physical system during a thermodynamic process is necessary for assessing the power and efficiency of thermal engines and refrigerators. In the last decade, quantum thermodynamic machines operating with a quantum working fluid have received renewed interest \cite{EspositoRMP,CampisiRMP,GooldReview,MillenXuereb,vinjanampathy2016quantum,pekola2015towards,KosloffOtto}. Several setups have been proposed for the realisation of thermal machines fully in the quantum regime and the first successful experiments have been reported~\cite{RossnagelScience,maslennikov2017}. 

\begin{figure}[t!]
\begin{center}
\includegraphics[width=0.99\columnwidth]{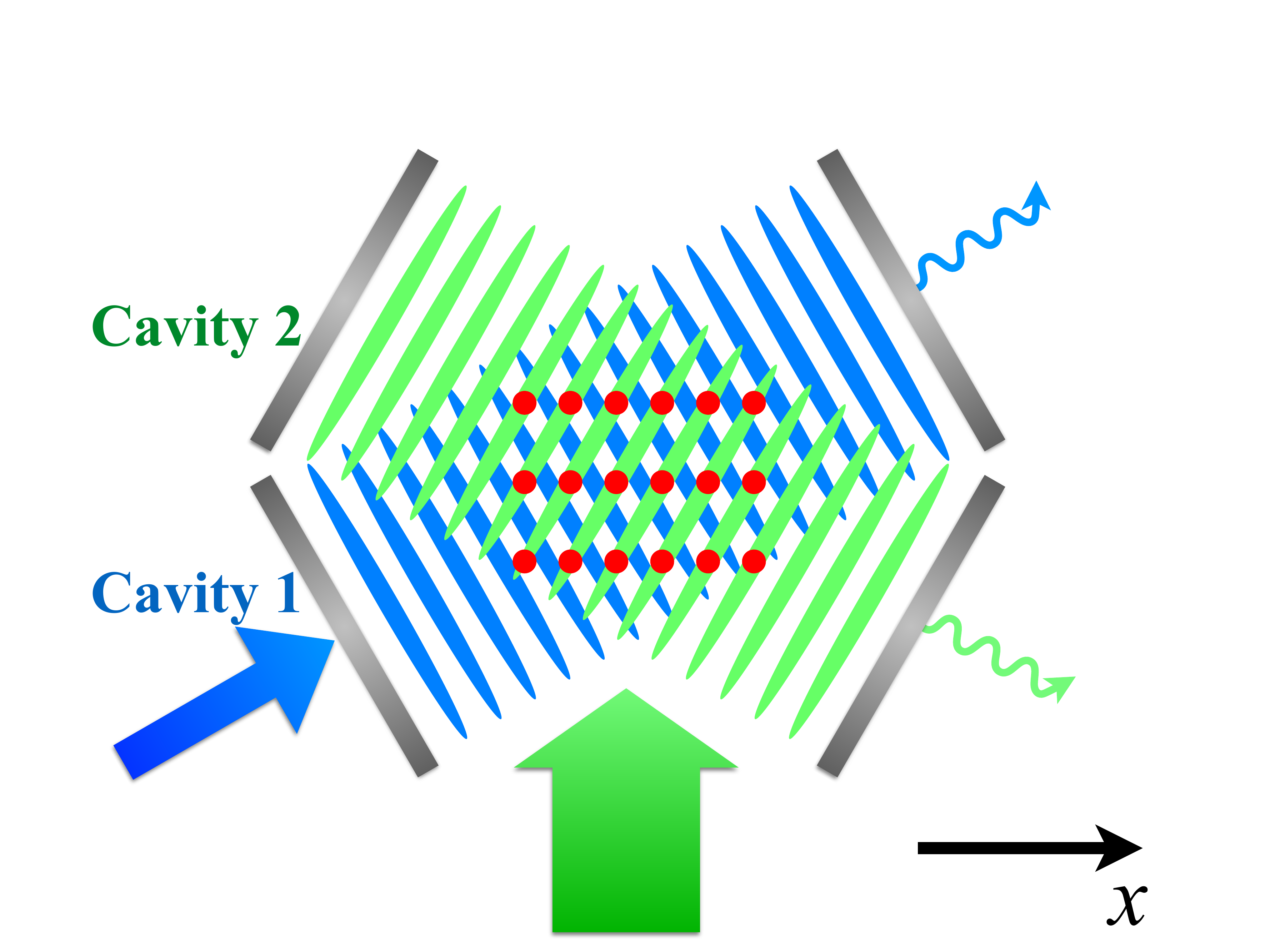}
\caption{Schematic setup for estimating the energy of lattice models. Individual atoms (red dots) are regularly arranged along one dimensional arrays by an external optical lattice parallel to the $x$ axis, not shown. The atoms interact with the fields of two optical cavities whose axis form an angle of 60 degrees as in \cite{Donner2017}. The atoms are accurately positioned at the nodes of the field of one of the two cavities (blue, labelled 1) which is externally pumped by a laser. Moreover the atoms are positioned at the antinodes of the other cavity field (green, labelled 2) and transversally pumped by an external laser. Tunneling processes will populate the blue cavity 1 and can be revealed by measuring the output quadrature. A photon number measurement of the green cavity 2 reveals instead the self-interaction term. We remark that our proposal is not restricted to this specific arrangement and that two overlapping but distinguishable optical modes would suffice.}
\label{fig:scheme}
\end{center}
\end{figure}

In order to assess the performance of such machines in terms of efficiency and output work power, one needs to monitor the energy of the quantum working substance. For strongly-interacting quantum systems, measuring the total energy is however challenging. In this paper, we show how to estimate the total internal energy of atoms in optical lattices by coupling the atoms to two optical cavities as recently achieved experimentally \cite{Donner2017} and shown schematically in Fig.~\ref{fig:scheme}.

For two-level systems, estimates of the work extracted and heat exchanged can be achieved by measuring the spin imbalance, by performing the full quantum state tomography~\cite{SerraMaxwellDemon} or by monitoring the system by a microwave cavity \cite{cottet2017}. For quantum harmonic oscillators embodied by trapped ions this can be achieved by measuring the phononic occupation \cite{an2015experimental,RossnagelScience,maslennikov2017} while for optomechanical systems work can be assessed by monitoring the intracavity field~\cite{DongMeystre2015}. In solid state devices, work and heat can be indirectly estimated by measuring charge currents or  the change of temperature of a thermal resistor due to absorption of microwave radiation \cite{Saira2012,PekolaCalorimetric}.

A Ramsey scheme which involves coupling the quantum working substance to an external two-level ancilla has been proposed for measuring work and, in general, internal energy changes \cite{Dorner2013,Mazzola2013,CampisicQED,GooldModi}. Its realisation in a nuclear magnetic resonant experiment allowed for the first experimental verification of the Jarzynski equality \cite{jarzynski1997nonequilibrium} in a quantum setting \cite{Batalhao}. Another scheme has been proposed which would be suitable for atomic ensembles and dimers in optical lattices and it involves coupling the atoms to the polarisation of a light mode \cite{DeChiaraPaz}. A simpler scheme adapted for atoms in double wells allows the measurement of the first moments of the work fluctuations just by measuring the population imbalance in the two wells and its fluctuations \cite{Lena}.

These methods are however unsuitable for quantum lattice gases. In fact, to employ the Ramsey scheme one would need to couple an ancilla to the whole lattice which seems unrealistic. The light assisted method presented in Ref.~\cite{DeChiaraPaz} does not resolve the individual atom-atom correlations accounting for all energetic terms in the system Hamiltonian. The recently developed single atom microscope allows for the precise measurement of atomic occupation of each lattice site~\cite{Bakr2010,Sherson2010} but currently not for the simultaneous measurement of atomic occupation and coherence that would be necessary for measuring the Hamiltonian operator. On the other hand, an approximate estimate of the energy change during a transformation, e.g. thermalisation, can be obtained in terms of local observables~\cite{Kaufman2016}.

Here, we propose an alternative route based on the strong interaction of atoms immersed in optical cavities which have been the subject of numerous investigations both theoretically \cite{MaschlerPRL2005,Larson2006,Mekhov2007,Sonia,Meystre2010,Mekhov2012,habibian2013a,habibian2013b,Mekhov2015,Zuppardo2015,Piazza2017} and, in recent years, experimentally~\cite{Brennecke2008,Murch2008,Landig}. 

Our proposal to estimate the total internal energy of atoms in optical lattices consists in coupling the atoms to two optical cavities as recently achieved experimentally \cite{Donner2017} and shown schematically in Fig.~\ref{fig:scheme}. We assume the atoms to be governed by the celebrated Bose-Hubbard Hamiltonian, described in Sec.~\ref{sec:model}, and in Sec.~\ref{sec:mapping} we explain how the tunnelling and self-interaction terms of the Hamiltonian can be mapped to the field operators of the two cavities in their steady state. As the cavities decay is typically much faster than the timescale of the atomic dynamics, our method would allow for the continuous monitoring of the atoms internal energy during a thermodynamic process. In Sec.~\ref{sec:numerical} we compare the ground state energy estimated with the cavity method with the exact ground state energy obtained numerically with density matrix renormalisation group simulations and find very good agreement. Finally, in Sec.~\ref{sec:conclusions} we summarise and conclude. In the next section we review the concepts of heat and work in quantum systems.

\section{Heat and work in quantum systems}
\label{sec:work}
For classical systems undergoing a transformation under  the influence of an external force and in contact to an external reservoir the first law of thermodynamics establishes the balance of the internal energy $\Delta U_{\rm int}$, the work $W$ done or extracted from the system and the heat $Q$ exchanged with the reservoir:
\begin{equation}
\Delta U_{\rm int} = Q+W
\end{equation}
where we have used the common convention that $Q>0$ represents heat absorbed by the system and $W<0$ is work extracted.

For quantum systems a similar relation holds provided that we identify the internal energy of the system with the mean value of the Hamiltonian $H_S$ governing its dynamics: $U_{\rm int} = {\rm Tr}(\rho H_S)$ where $\rho$ is the density matrix of the system. If the system undergoes a transformation that changes its Hamiltonian from $H_S(0)$ at time $t=0$ to $H_S(\tau)$ at the final time $t=\tau$ then we have:
\begin{equation}
\dot U_{\rm int}=   {\rm Tr}[\dot\rho(t) H_S(t)]+ {\rm Tr}[\rho(t) \dot H_S(t)].
\end{equation}
On the right hand side of this equation, the first term represents the heat current $\dot Q$ while the second term is the work power $\dot W$.
Integrating the power we obtain a discrete equation for the net work:
\begin{equation}
\label{eq:averagework}
W={\rm Tr}[\rho(\tau) H_S(\tau)] - {\rm Tr}[\rho(0) H_S(0)]. 
\end{equation}
This definition of the net work is compatible with the so-called two-point energy measurement definition of work in quantum mechanics which consists in measuring the system energy at time $t=0$, performing the transformation from $t=0$ to $t=\tau$, and finally measuring again the system energy at $t=\tau$ \cite{LutzPRE2007}. Work thus becomes a stochastic variable subject to intrinsic quantum fluctuations of the Hamiltonian operator. Similarly the heat exchanged can in principle be estimated by measuring the energy balance of the reservoir if it is accessible.

Although the two-point measurement is the most common definition of work in the quantum community, having the advantage of fulfilling the first law and the Jarzynski equality \cite{jarzynski1997nonequilibrium}, it is not the only one. In fact it has been criticised because it does not take into account the work performed during the initial energy measurement which collapses the system state onto one eigenstate of the Hamiltonian destroying all initial coherences. As a result, other definitions have been recently proposed \cite{Gallego2016,TalknerHanggiPRE2016,Acin2017, miller_anders}.
 In most of these proposals however a central requirement is that the energy of the system is measured. In this work, regardless of the definition of work, we propose how to measure the energy of a system of atoms in optical lattices. 

\section{Model}
\label{sec:model}

\subsection{Atom-photon interactions}
In this section we briefly revise the physics of non-interacting two-level atoms interacting with the electromagnetic field of one or more optical resonators~\cite{walls2007quantum}. We assume the atoms to be tightly confined in one-dimensional (1D) tubes parallel to the $x$ axis and subject to an optical lattice described by a periodic potential $V(x)$ with periodicity $d$ (see Fig.~\ref{fig:scheme}).  We stress that the potential $V(x)$ is not generated by the fields of the cavities and can be realised with an external optical field. This could be generated by a single atom microscope \cite{Bakr2010,Sherson2010} which would ensure both trapping and precise positioning of the atoms in the field of the optical cavities.

We also assume the atoms to be pumped by a laser detuned by $\Delta_a$ from the atomic transition and let us define $\sigma~ (\sigma^\dagger)$ the lowering (raising) operator of the corresponding transition.
Assuming the dipole and rotating wave approximation the total Hamiltonian for a single atom reads: 
\begin{eqnarray}
h&=& h_A+H_C-\hbar \Delta_a\sigma^\dagger\sigma
+\hbar\Omega_0(\sigma+\sigma^\dagger) \\
&+&\hbar\sum_{l=1}^{N_c} g_{l}(x)\left(\sigma^\dagger a_l+\sigma a_l^\dagger\right)
\end{eqnarray}
where $\Omega_0$ is the atomic Rabi frequency, $g_l(x)=g_l u_l(x)$ is the atom-cavity field coupling strength proportional to the real spatial profile $u_l(x)$ of the cavity field at position $x$ and where
\begin{eqnarray}
h_A=\frac{p^2}{2m} +V(x)
\end{eqnarray}
is the Hamiltonian of the spatial degree of freedom of the atom;
the term
\begin{eqnarray}
H_C=-\hbar\sum_{l=1}^{N_c}  \Delta_{cl} a_l^\dagger a_l +\hbar\sum_{l=1}^{N_c}\eta_l (a_l+a^\dagger_l)
\end{eqnarray}
is the Hamiltonian for the $N_c$ cavities (or different modes of the same cavity) where $a_l$ ($a^\dagger_l$) is the annihilation (creation) operator of the $l$th cavity with detuning $\Delta_{cl}$ and $\eta_l$ is the pump strength for each cavity. We assume that the atomic Rabi frequency $\Omega_0$ does not strongly depend on $x$, i.e. the spatial variation of the transverse atomic pump is negligible in the region in which the atoms are illuminated.

Assuming large atomic detuning from the excited level, the population of the excited state is small and we can adiabatically eliminate it. Under these approximations, the complete Hamiltonian of the atom-cavity interaction becomes: $h=h_A+H_C+h_{AC}$ where
\begin{eqnarray}
\label{eq:HAC}
h_{AC}&=& \sum_{l=1}^{N_c}\frac{\hbar\Omega_0 g_l(x)}{\Delta_a}(a_l+a^\dagger_l)+
\\
&+&\sum_{l,m=1}^{N_c} \frac{\hbar g_l(x)g_m(x)}{\Delta_a} a_l^\dagger a_m 
\nonumber
\end{eqnarray}
is the renormalised atom-cavities interaction Hamiltonian in which we have neglected constant terms.

\subsection{Many-Body Hamiltonian}
In the more general case of many interacting bosons we use second quantization to write the system Hamiltonian by introducing the bosonic atomic field operator $\psi(x)$ that destroys an atom at position $x$. 
The derivation of the effective Hamiltonian follows various references \cite{MaschlerPRL2005,Larson2006,Mekhov2007,Sonia,Meystre2010,Mekhov2012,Mekhov2015}. We assume that atoms are ultracold so that they only interact via s-wave scattering with scattering length $a_s$. 
The many-body Hamiltonian can therefore be expressed as:
\begin{eqnarray}
H&=&H_C+\int\mathrm{d}x~\psi^{\dagger}(x)h_A
{\psi}(x)
\nonumber
\\
&+&\dfrac{2\pi\hbar^{2}a_{s}}{m}\int\mathrm{d}x~ {\psi}^{\dagger}(x) {\psi}^{\dagger}(x) {\psi}(x) {\psi}(x)
\label{eq:hamiltonian_general_second_quantization}
\\
&+&H_{AC} 
\nonumber
\end{eqnarray}
where $H_{AC} = \int\mathrm{d}x~ {\psi}^{\dagger}(x)h_{AC}{\psi}(x)$ is the second quantised version of the atoms-cavity interaction.

Within the tight binding and single band approximations \cite{cold}, we can thus expand the atomic field operator in terms of real orthonormal Wannier functions localised at the minima $x_i = id$ of the trapping potential:
\begin{equation}
\label{eq:psi}
\psi(x) = \sum_{i=1}^M w(x-x_i) b_i
\end{equation}
where the operators $b_i$ are the bosonic operators that annihilate an atom in site $i$ and satisfy bosonic commutation relations $[b_i,b^\dagger_j]=\delta_{ij}$ and $M$ is the number of sites. 
Similar to Ref.~\cite{Mekhov2015}, when we insert Eq.~\eqref{eq:psi} in the many-body Hamiltonian Eq.~\eqref{eq:hamiltonian_general_second_quantization} the following overlap integrals coefficients appear:
\begin{eqnarray}
\label{eq:Jcl}
J_{i,j}^{\text{cl}}&=&\int\mathrm{d}x~w(x-x_{i})h_A
w(x-x_{j})\\
J_{i,j}^{lm}&=&\int\mathrm{d}x~w(x-x_{i})u_{l}(x)u_{m}(x)w(x-x_{j})\nonumber\\
\label{eq:Jlm}
&&\\
U&=&\frac{4\pi\hbar^2a_s}{m} \int\mathrm{d}x~ w^4(x)
\end{eqnarray}
The coefficients $J_{i,j}^{\text{cl}}$ in Eq.~\eqref{eq:Jcl} arise from the atomic kinetic energy and the interaction of the atoms with the ``classical" potential $V(x)$ that is not generated by the cavities and are typical in the derivation of the Bose-Hubbard model. On the other hand the coefficients $J_{i,j}^{lm}$ are responsible for tunnelling terms and atomic shifts assisted by photon absorption/emission into the cavity modes $l$ and $m$. For convenience we also define in Eq.~\eqref{eq:Jlm} the coefficients $J_{i,j}^{0l}$ by setting $u_0(x)=1$. This will be useful for defining overlap integrals involving the cavity mode $l$ and the external laser pump assumed with no spatial variation along $x$. For a deep lattice potential $V(x)$ we can assume that the overlap integrals are nonzero only for onsite terms $(i=j)$ or for nearest-neighbor terms $|i-j|=1$ and we obtain the following Hamiltonian:
\begin{equation}
{H}=H_{C}+{H}_{\text{BH}}+H_{AC}
\end{equation}
where 
\begin{equation}
H_{\text{BH}}=-J\sum_{i=1}^{M-1}({b}^{\dagger}_{i}{b}_{i+1}+h.c.)+\frac{U}{2}\sum\limits_{i=1}^{M}{n}_{i}({n}_{i}-1)
\end{equation}
is the Bose-Hubbard Hamiltonian where we have set $J=J_{i,i\pm1}^{\text{cl}}$ and we are assuming open boundary conditions. The atom-cavities interaction Hamiltonian $H_{AC}$ reported in Eq.~\eqref{eq:HAC} for a single atom, in second quantization takes the form
\begin{eqnarray*}
\label{eq:HAC2}
H_{AC}&=&\sum_{l=1}^{N_c}\frac{\hbar\Omega_{0}g_{l}}{\Delta_{a}}\left(a^\dagger_{l}+{a}_{l}\right){\mathcal{F}}_{0l}
\\
&+&\sum\limits_{l,m=1}^{N_c}\frac{\hbar g_{m}g_{l}}{\Delta_{a}}{a}^{\dagger}_{l}{a}_{m}{\mathcal{F}}_{lm}
\end{eqnarray*}
where we have introduced the atomic operators ${\mathcal{F}}_{lm}={\mathcal{D}}_{lm}+{\mathcal{B}}_{lm}$ such that:
\begin{eqnarray}
{\mathcal{D}}_{lm}&=&\sum\limits_{i=1}^{M}J_{i,i}^{lm}{n}_{i},\\
{\mathcal{B}}_{lm}&=&\sum\limits_{i=1}^{M-1}J_{i,i+1}^{lm}({b}_{i}^{\dagger}{b}_{i+1}+h.c.)
\end{eqnarray}
Notice that while the operator ${\mathcal{D}}_{lm}$ depends on the atomic spatial distribution, the operator ${\mathcal{B}}_{lm}$ measures the spatial coherence. In the next section we explain how these operators can be indirectly measured by observing the cavities output light (see also \cite{Mekhov2015} in a different context).

\subsection{Probing the atomic properties using the cavity output light}
Let us write the Heisenberg-Langevin equations for the cavities annihilation operators:
\begin{eqnarray}
\dot a_l &=& i[H,a_l]-\kappa a_l +\sqrt{2\kappa}\,a_l^{\rm in}(t)
\nonumber\\ 
&=&(i\Delta_{cl}-\kappa)a_l -i\eta_l -i\frac{\Omega_{0}g_{l}}{\Delta_{a}}{\mathcal{F}}_{0l}
\nonumber\\
\label{eq:Langevin}
&-&\sum\limits_{m=1}^{N_c}\frac{ig_{m}g_{l}}{\Delta_{a}}{a}_{m}{\mathcal{F}}_{lm}+\sqrt{2\kappa}\,a_l^{\rm in}(t)
\end{eqnarray}
where $\kappa$ is the cavity decay rate and the cavity noise operators $a_l^{\rm in}(t)$  are Langevin forces with zero mean and delta correlations: $\langle a_l^{\rm in}(t){a_m^{\rm in}}^\dagger(t') \rangle =\delta_{lm}\delta(t-t')$, see for instance \cite{carmichael}. In the limit of large cavity detuning $\Delta_{cl}$, the dynamics of the cavity is much faster than the dynamics of the atoms. Thus it is possible to find the steady state solution for the cavity field operators in Eq.~\eqref{eq:Langevin} in terms of the atomic operators $\mathcal F_{lm}$. In the simplest case of only one cavity mode $l=1$ we obtain:
\begin{equation}
\label{eq:a1}
a_1 = \frac{i\eta_1+i\frac{\Omega_{0}g_{1}}{\Delta_{a}}{\mathcal{F}}_{01}-\sqrt{2\kappa}\,a_1^{\rm in}(t)}{i\Delta_{c1}-\kappa-i\frac{g_1^2}{\Delta_a}\mathcal F_{11}}
\end{equation}
 Eq.~\eqref{eq:a1} relates the steady state cavity operator $a_1$ with the pump strength $\eta_1$, with the cavity detuning $\Delta_{c1}$, its decay constant $\kappa$ and the input noise operator $a_1^{\rm in}(t)$. More importantly, it connects $a_1$ with two atomic operators ${\mathcal{F}}_{01}$ and ${\mathcal{F}}_{11}$, defined above. Thus, from the analysis of the output light statistics one can infer some of the atomic properties.
In fact, the two operators are both a linear combination of the modulated density $\mathcal D$ and coherence operator $\mathcal B$.
In the next section we show that, using two cavities arranged as in Fig.~\ref{fig:scheme}, we can measure the energy of the interacting atoms from the analysis of the output signal from the two cavities.

\section{Mapping the Bose-Hubbard Hamiltonian onto a light mode}
\label{sec:mapping}

In this section, we show how to map the Bose-Hubbard Hamiltonian onto the combination of the light fields emerging from two optical cavities. 
First of all we start by rewriting the Bose-Hubbard Hamiltonian as \cite{cold}:
\begin{equation}
\label{eq:bose_hubbard_hamiltonian_with_P}
{H}_{\text{BH}}=-JB+\frac{U}{2}P-\frac{U}{2}N
\end{equation}
where 
\begin{equation}
B=\sum_{i=1}^{M-1}({b}_{i}^{\dagger}{b}_{i+1}+h.c.),\quad
P=\sum_{i=1}^M {n}_{i}^{2},\quad
N=\sum_{i=1}^M {n}_{i},\quad
\end{equation}
The total number of particles $N$ is conserved as it commutes with the Hamiltonian: $[{H}_{\text{BH}},N]=0$. As this is not a dynamical variable of the system we assume that it is estimated during the preparation of the atomic gas in the setup and does not change significantly during the duration of the process. In Sec.~\ref{sec:numerical} we numerically analyse the effect of systematic errors in the estimate of $N$ on our measurement of the internal energy.
We also assume to accurately know the tunneling and self-interaction coefficients $J$ and $U$ from an independent measurement. These coefficients can be estimated indirectly form the trapping potential depth $V(x)$ and by performing numerical simulations to determine the band structure and Wannier functions. 

Our strategy for mapping the Hamiltonian ${H}_{\text{BH}}$ onto a light quadrature, inspired by Ref.~\cite{Mekhov2015}, consists in mapping separately the operators $B$ and $P$ onto the fields of two different cavities (or two cavity modes of the same cavity) so that $N_c=2$. 
We assume that the two modes have either different polarisation or  different frequency. This means that the cross terms $J_{i,j}^{12}$ are all vanishing and thus that each cavity field is not directly influenced by the other cavity. In other words, we are neglecting scattering processes in which one photon is absorbed by the atoms from one cavity and, as a result, re-emitted into the other cavity. 

To map the tunneling operator $B$ onto the field of the first cavity ($l=1$), we assume that the cavity is pumped by an external laser with amplitude $\eta_1$ and the atoms are not directly pumped by an external laser with this polarisation or frequency. Therefore the overlap integral $J^{01}_{i,j}=0$ and as a consequence $\mathcal F_{01}=0$. We choose the spatial mode of the first cavity to be:
\begin{equation}
u_1(x) = \sin\left(\frac{\pi x}{d}\right )
\end{equation}
where, as before, $d$ is the inter-atomic distance. 
The steady state output quadrature of the cavity becomes:
\begin{eqnarray}
\label{eq:quadrature_measured_first_polarization}
\langle {a}_{1}+{a}_{1}^{\dagger}\rangle
&\simeq&
\dfrac{2\eta_{1}}{\kappa^{2}+\Delta_{c1}^{2}}\left(\Delta_{c1}-\dfrac{g_{1}^{2}{\langle \mathcal{F} }_{11}\rangle}{\Delta_{a1}}\right)
\end{eqnarray}
where in the last expression we retained only first order contributions of $\langle \mathcal F_{11}\rangle$ assuming $g_1^2/\Delta_{a1} \ll \kappa,\Delta_{c1}$.
In the tight binding regime, Wannier functions are well localised and only a small overlap between nearest neighbors exists. In this regime the expression of the atomic operator $\mathcal F_{11}$ becomes \cite{Mekhov2015}: $\mathcal F_{11} = J^{11}_{ii} N + J^{11}_{i,i+1} B$.
In practice $J^{11}_{i,i+1}\ll J^{11}_{ii}$ however the first contribution is proportional to the total number of particles and does not induce quantum fluctuations. Moreover, from a measurement of the number of particles, this term can be estimated. To summarize this step: by measuring the output cavity quadrature of the $l=1$ cavity we can measure the mean value of the tunneling operator $B$ of the Bose-Hubbard model:
\begin{equation}
\label{eq:a1B}
\langle a_1+a_1^\dagger \rangle=\chi_0+\chi_1 N +\chi_2 \langle B \rangle
\end{equation}
where we defined 
\begin{eqnarray*}
\chi_0&=&(2\eta_1\Delta_{c1})/(\kappa^2+\Delta_{c1}^2)
\\
\chi_1&=&-(2\eta_1g_1^2/\Delta_{a1})J_{i,i}^{11}/(\kappa^2+\Delta_{c1}^2)
\\
\chi_2&=&-(2\eta_1g_1^2/\Delta_{a1})J_{i,i+1}^{11}/(\kappa^2+\Delta_{c1}^2).
\end{eqnarray*}
We note that by pumping this cavity with an external laser, the contribution of $\langle B \rangle$ to $\langle a_1+a_1^\dagger \rangle$ can be further enhanced as described in Ref.~\cite{Mekhov2015}.

We would like to stress that the measurement of the atomic field coherence $\langle B \rangle$ can be interpreted as a dispersive shift caused by the atoms that act as a dielectric medium inside the cavity and can therefore be revealed by analysing the cavity output light using standard optical measurements.

Now we turn to the self-interaction term $P$ of the Bose-Hubbard Hamiltonian. This depends on the atomic density squared in each site and, as such, it cannot be easily measured with a light-assisted measurement. To overcome this problem we proceed as follows. We consider the second cavity $l=2$ without external laser pump ($\eta_2=0$), whose cavity mode overlap with the atoms that are pumped with a laser of Rabi frequency $\Omega_0$ with the same polarisation as the cavity mode (see Fig.~\ref{fig:scheme}). This means that the cavity field is built by the photons that are absorbed from the laser by the atoms and emitted into the cavity. In this case the steady state output cavity field is:
\begin{eqnarray*}
{a}_{2}&=&\frac{\Omega_{0}g_{2}{\mathcal{F}}_{20} -\sqrt{2\kappa}a_2^{\rm in}(t)}{\Delta_{a2}\left(\Delta_{c2}+i\kappa\right)-g_{2}^{2}{\mathcal{F}}_{22}}
\\
&\simeq& \frac{\Omega_{0}g_{2}{\mathcal{F}}_{20}-\sqrt{2\kappa}a_2^{\rm in}(t)}{\Delta_{a2}\left(\Delta_{c2}+i\kappa\right)}
\end{eqnarray*}
where in the last expression we have neglected shifts in the effective cavity detuning due to the atomic distribution. If we choose the mode cavity to be $u_2(x)=\cos(\pi x/d)$, then the overlap integrals become:
\begin{equation}
J^{20}_{i,i} =(-1)^i J^{20};\quad J^{20}_{i,i+1}=0
\end{equation}
because of symmetry reasons. The actual value of $J^{20}$ given by the overlap integral of Eq.~\eqref{eq:Jlm} depends on the specific form of the Wannier functions and can be evaluated numerically. Thus the number of photons emitted from the second cavity is:
\begin{eqnarray}
\label{eq:a2a2}
\langle a_2^\dagger a_2\rangle &=&|R|^2 \langle \mathcal D_{20}^\dagger \mathcal D_{20}\rangle=\\
&=& |R|^2(J^{20})^2 \sum_{i,j=1}^M (-1)^{i+j} \langle n_i n_j \rangle
\nonumber \\
&=& |R|^2(J^{20})^2 \left[P + 2\sum_{i<j}^M (-1)^{i+j} \langle n_i n_j \rangle \right]
\nonumber
\end{eqnarray}
where we have set
\begin{equation}
R= \frac{\Omega_0 g_2}{\Delta_{a2}(\Delta_{c2}+i\kappa)}
\end{equation}

Now we also have:
\begin{equation}
\label{eq:N2}
N^2= P +2\sum_{i<j}^M \langle n_i n_j \rangle
\end{equation}
and summing it to Eq.~\eqref{eq:a2a2} we obtain:
\begin{widetext}
\begin{equation}
\label{eq:a2a22}
\langle a_2^\dagger a_2\rangle = |R|^2(J^{20})^2 \left\{ 2P-N^2+ 2\sum_{i<j}^M [1+(-1)^{i+j}] \langle n_i n_j \rangle \right\}.
\end{equation}
\end{widetext}
This expression contains two-point density correlations $\langle n_i n_j \rangle$. However, notice that nearest-neihbor correlations $\langle n_i n_{i+1} \rangle$ do not enter the sum thanks to the $(-1)^{2i+1}$ factor. In the Mott insulator phase, by neglecting corrections that decay exponentially with the distance $|i-j|$, the correlations $\langle n_i n_j \rangle$ can be approximated by the product of expectation values $\langle n_i\rangle \langle n_j \rangle=n^2$, where $n=N/M$ is the average uniform filling. This is a strong approximation which we verify numerically in Sec.~\ref{sec:numerical}. 
Under these assumptions we finally obtain:
\begin{equation}
\label{eq:a22}
\langle a_2^\dagger a_2\rangle \simeq \xi P+\alpha
\end{equation}
where $\xi=2|R|^2(J^{20})^2$ and
$$
\alpha= |R|^2(J^{20})^2\left\{ -N^2+ 4n^2 \left(\frac M2-1\right)\frac M2 \right\}.
 $$
Finally, by putting together the mean values of Eqs.~\eqref{eq:quadrature_measured_first_polarization} and \eqref{eq:a22}, we obtain that the expectation value of the operator
\begin{equation}
\label{eq:Q}
\langle G \rangle= -\frac{J}{\chi_2} \langle a_1+a_1^\dagger\rangle +\frac{U}{2\xi}\langle a_2^\dagger a_2\rangle -E_0 \simeq\langle H_{BH} \rangle
\end{equation}
is proportional to the mean energy value of the Bose-Hubbard Hamiltonian. 
In Eq.~\eqref{eq:Q} we have set:
\begin{equation}
E_{0} = \frac{UN}{2}+\frac{J(\chi_0+\chi_1N)}{\chi_2} +\frac{\alpha U}{2\xi}
\end{equation}
 which is a constant term not affected by quantum fluctuations and that can be estimated through the experimental parameters of the cavities and the optical lattice trapping the atoms.

Eq.~\eqref{eq:Q} is the main result of the paper. The approximated relation $\langle G\rangle \simeq\langle H_{BH} \rangle$ originates from the replacement of Eq.~\eqref{eq:a2a22} with Eq.~\eqref{eq:a22} which is valid for short range correlations $\langle n_i n_j \rangle$. This approximation is quite accurate in the Mott insulator phase ($J\ll U$) and becomes worse and worse as $J$ increases. A numerical assessment of the quality of this mapping is presented in the next section.

\section{Numerical results}
\label{sec:numerical}

In this section we compare the energy estimated with the technique developed in the previous section with the exact ground state energy computed numerically. We use the density matrix renormalisation group (DMRG) algorithm \cite{White92,SchollwockRMP,DeChiara08} with open boundary conditions to calculate the ground state energy. We consider lengths $M=40,80$ which are not much larger than the sizes reachable in current experiments. We also added an uncertainty in the cavity estimated values assuming an error of 10\% in the estimate of the total number of particles $N$.

\begin{figure}[t!]
\begin{center}
\includegraphics[width=0.99\columnwidth]{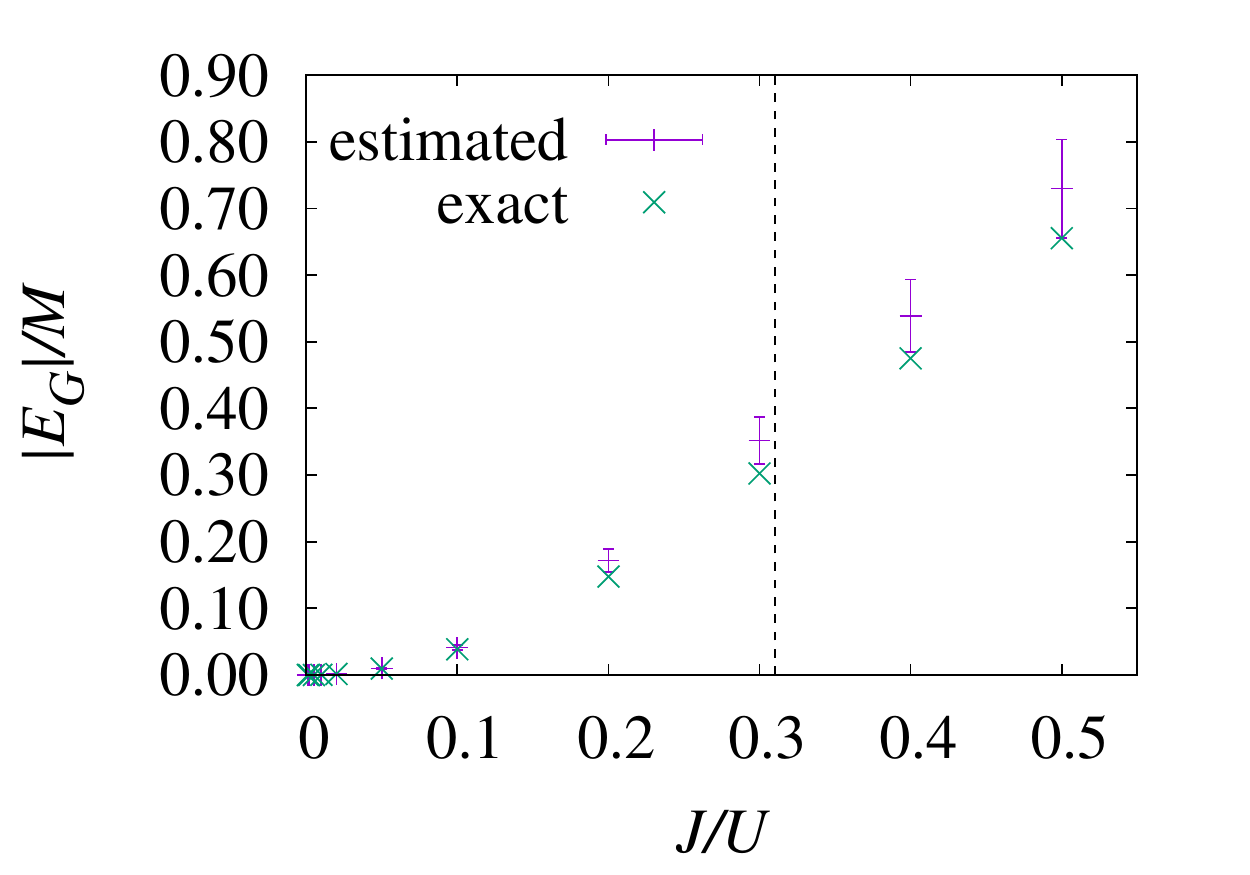}
\includegraphics[width=0.99\columnwidth]{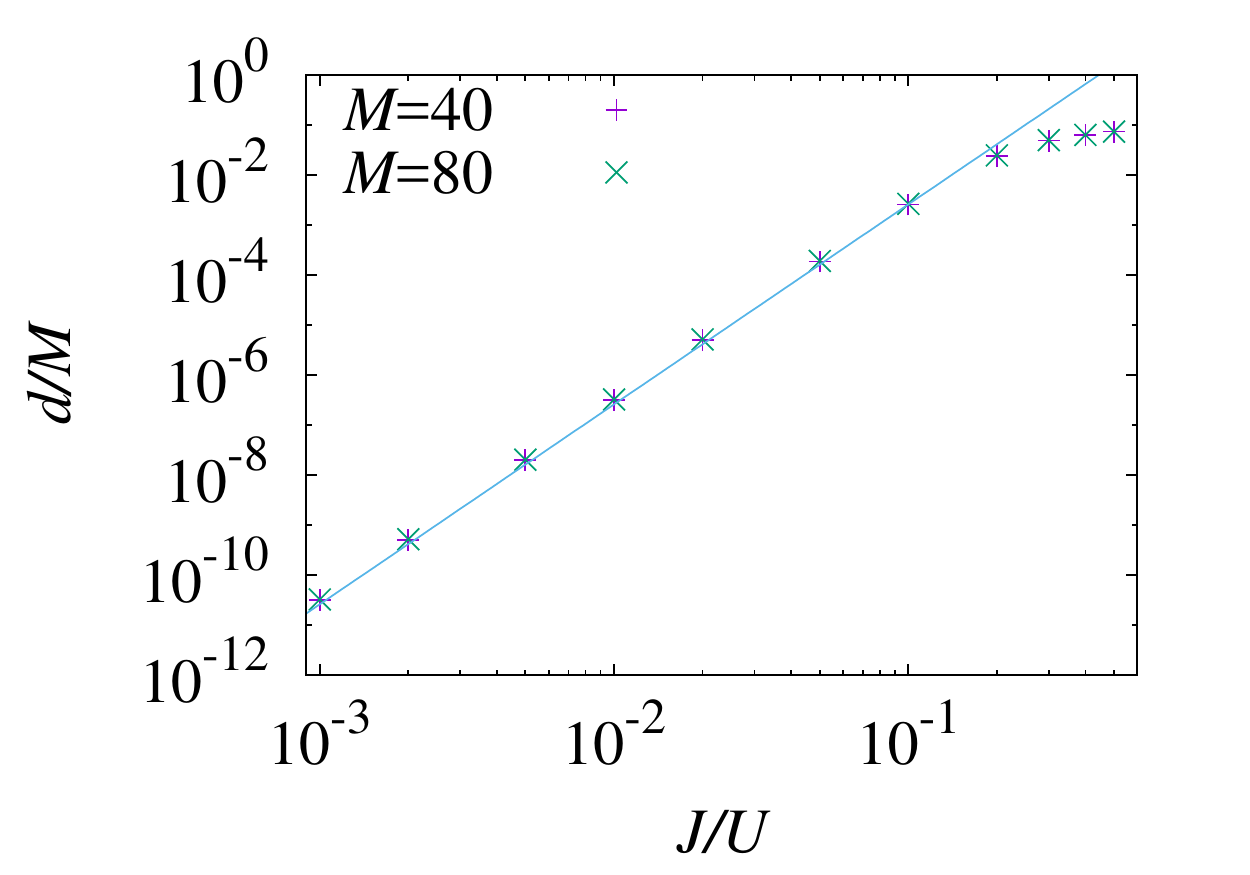}
\caption{(a) Comparison of the exact ground state energy ($\times$) with the value estimated using the cavity method (+) for $M,N=40$ against $J/U$. The error bars represent the interval of uncertainty when a 10\% fluctuation is added to the estimated values. A vertical dashed line at $J=0.31 U$ indicates the approximate location of the Mott insulator to superfluid transition. (b) Absolute difference $d$ between the exact ground state energy and the estimated one against $J/U$ in double logarithmic scale for $M,N=40,80$. The solid line is a power law fitting $d\propto (J/U)^4$. All energies are per particle.}
\label{fig:comparison}
\end{center}
\end{figure}

The results are shown in Fig.~\ref{fig:comparison}. In panel (a) we show the ground state energy $|E_G|/M$ per particle for $M=40$ sites with unit filling $N=M$. The estimated and exact values are very close for $J\ll U$ and their absolute difference starts being significant for $J\simeq 0.2 U$ which is quite close to the Mott Insulator to superfluid transition $J\simeq 0.31 U$ indicated in the plot by a vertical dashed line. The reason of this discrepancy lies in our approximation of replacing the correlation $\langle n_i n_j \rangle$ with the square of the filling $n$ for further than nearest-neighbors.

In panel (b) we show the difference $d$ of the exact ground state energy and the estimated one per particle for $M=40,80$. From the plot it is evident that there is no significant dependence on the number of particles. Moreover for small $J/U$ the discrepancy grows approximately as a power law:
\begin{equation}
d \sim \left(\frac JU\right ) ^4
\end{equation}
which can be understood using a strong coupling expansion argument \cite{Monien}. There, it was shown that the lowest ground state energy correction is of second order in $J/U$ while the third order vanishes. This is due to virtual processes in which a boson hops to the nearest neighbour site and back. However thanks to the trick of summing Eq.~\eqref{eq:N2} to Eq.~\eqref{eq:a2a22} we have eliminated such nearest-neighbour correction. Thus, the next non zero term is a fourth order hopping virtual process in which a boson hops to the next-to-nearest neighbour and back. 
For larger values of $J/U$ the discrepancy grows much slower.

Using our technique we can estimate the work done on or extracted from the atoms when one parameter of the atomic Hamiltonian changes. As an  example, we here discuss the case of an instantaneous quench which has been treated very often in the literature~\cite{Silva,Dorner2012,Marino,Yulia,Sindona,Masca,Fusco,Zhong,Bayocboc,Paganelli,Cosco}. We assume that the tunnelling coefficient is changed instantaneously from $J$ to $J+\Delta J$. From Eq.~\eqref{eq:averagework} we find:  
\begin{equation}
W=-\Delta J \langle B\rangle
\end{equation}
where the average is taken over the initial state. From Eq.~\eqref{eq:a1B}, the expectation value of the tunnelling operator $\langle B\rangle$ can be exactly measured through one of the quadratures of cavity 1 (assuming that the parameters $\chi_i$ and the number of particles $N$ can be accurately estimated). 

\begin{figure}[t]
\begin{center}
\includegraphics[width=0.9\columnwidth]{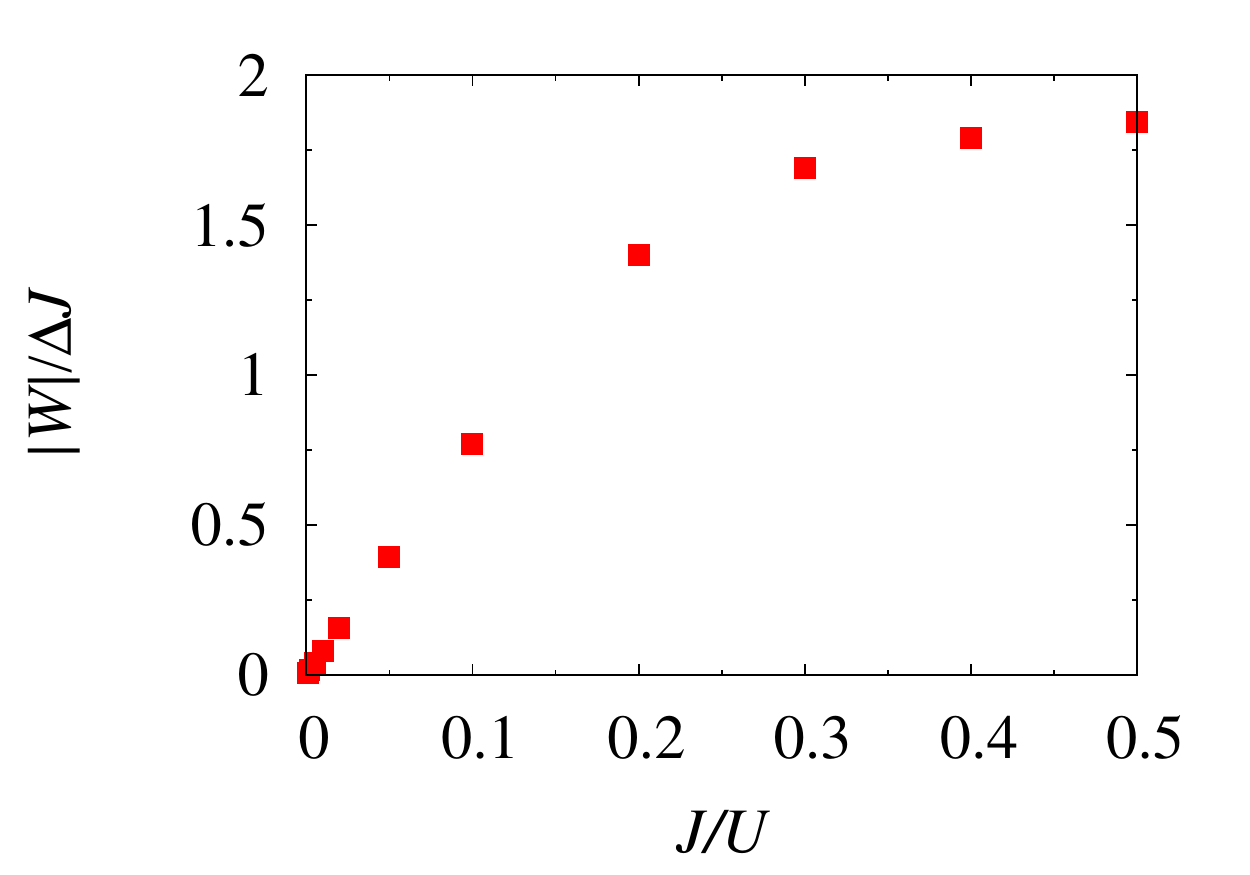}
\caption{Absolute value of the work done/extracted after an instantaneous quench of the tunnelling from $J$ to $J+\Delta J$ in units of the tunnelling increment $\Delta J$ as a function of the initial value of the tunnelling $J$. We choose $M=80$ and unit filling.}
\label{fig:work}
\end{center}
\end{figure}

We calculated numerically the value of the average work using DMRG for a system of interacting bosons described by the Bose-Hubbard Hamiltonian $H_{BH}$ with $M=80$ sites and unit filling. The results, plotted in Fig.~\ref{fig:work} show how the ratio $|W|/\Delta J$ increases before saturating for $J$ larger than the critical point. In the superfluid phase, since the state does not change significantly, the amount of work done/extracted is simply proportional to $\Delta J$.

\section{Conclusion}
\label{sec:conclusions}

Summarising, in this paper we have shown how to measure the internal energy of a system made of ultracold atoms trapped in a one-dimensional optical lattice. The atomic Hamiltonian operator is mapped to two operators of the fields of two optical cavities which can then be revealed with standard quantum optical measurements. Our result can be used for the monitoring of internal energy, work extracted and heat currents in optical lattice gases employed as working fluids for quantum thermal machines. 

The setup shown in Fig.~\ref{fig:scheme} is inspired by the experiment reported in Ref.~\cite{Donner2017} in which the two cavities form an angle of 60 degrees. We remark, however, that our proposal is much more general: two cavities crossed at an arbitrary angle would suffice. Alternatively our proposal could be realised with a single cavity and two modes with different polarisation or different periodicity.

It is important to stress that in our derivation we need to assume the form of the Hamiltonian and the knowledge of the tunnelling coefficient $J$ and self-interaction constant $U$. However, our method is general and can be extended to the fermionic Hubbard model and to spinor Bose- and Fermi-Hubbard models.

\acknowledgements
The authors thank T. Donner and I. Mekhov for invaluable discussions and for comments on the manuscript. 
LV acknowledges the kind hospitality at the CTAMOP centre and the Quantum technology group at Queen's University Belfast.

\bibliographystyle{apsrev4-1}
\bibliography{biblio}


\end{document}